# Evaluating the accuracy of diffusion MRI models in white matter

*Ariel Rokem[1*], Jason D. Yeatman[1,2], Franco Pestilli[1,3], Kendrick N. Kay[1,4], Aviv Mezer[1,5], Stefan van der Walt[6], and Brian A. Wandell[1]*

1. Department of Psychology, Stanford, Stanford 94305, CA, USA
2. Institute for Learning and Brain Sciences, University of Washington, Seattle, WA 98195, USA
3. Department of Psychological and Brain Sciences, Program in Neuroscience, Indiana University, Bloomington, IN 47405, USA
4. Department of Psychology, Washington University in St. Louis, St. Louis, MO, 63130, USA
5. Edmond and Lily Safra Center for Brain Sciences (ELSC), The Hebrew University, Givat Ram, Jerusalem 91904, Israel
6. Division of Applied Mathematics, Stellenbosch University, Stellenbosch 7600, South Africa.
*Corresponding author: arokem@gmail.com

# Abstract

Models of diffusion MRI within a voxel are useful for making inferences about the properties of the tissue and inferring fiber orientation distribution used by tractography algorithms. A useful model must fit the data accurately. However, evaluations of model-accuracy of commonly used models have not been published before. Here, we evaluate model-accuracy of the two main classes of diffusion MRI models. The diffusion tensor model (DTM) summarizes diffusion as a 3-dimensional Gaussian distribution. Sparse fascicle models (SFM) summarize the signal as a sum of signals originating from a collection of fascicles oriented in different directions. We use cross-validation to assess model-accuracy at different gradient amplitudes (b-values) throughout the white matter. Specifically, we fit each model to all the white matter voxels in one data set and then use the model to predict a second, independent data set. This is the first evaluation of model-accuracy of these models. In most of the white matter the DTM predicts the data more accurately than test-retest reliability; SFM model-accuracy is higher than test-retest reliability and also higher than the DTM model-accuracy, particularly for measurements with (a) a b-value above 1000 in locations containing fiber crossings, and (b) in the regions of the brain surrounding the optic radiations. The SFM also has better parameter-validity: it more accurately estimates the fiber orientation distribution function (fODF) in each voxel, which is useful for fiber tracking.

# Introduction

Diffusion-weighted imaging (DWI) using MR has enormously expanded our understanding of the structures and connections in the living human brain. The interest in this technology has given rise to a wide array of efforts to model the DWI signals. The purpose of these models is to clarify the biological structures that determine the signal. Based on these models, investigators make inferences about local tissue properties such as the orientation [1–3] coherence [4,5], and axon size-distribution [6,7] of white-matter fiber bundles (or fascicles).

## Model evaluation

There are a large number of models of the diffusion signals measured within a voxel (reviewed in [8]), and there are several different approaches to assessing the value of these models. In one approach, investigators assess whether the model parameters provide useful information about specific aspects of the underlying biological tissue (parameter-validity). Parameter-validity is assessed by comparing parameter estimates with known anatomy, or by using phantoms constructed with specific parameters.

Models can also be evaluated by measuring parameter-reliability. One way to assess parameter-reliability is to compare the estimates across plausible noise levels, say the noise that arises across repeated measurements. A second way is to measure the effect of changes in the MR acquisition parameters. A substantial literature examines the parameter-reliability of common diffusion models [9–12] and particularly for differences in measurement parameters, such as the number of diffusion-weighting directions [13,10].

A third evaluation asks how accurately the model fits the measured signal (model-accuracy). Surprisingly, this aspect of the models has not been assessed extensively before (see Table 1). Model-accuracy differs from both parameter-validity and parameter-reliability. For example, parameter-reliability can be very high, but the model-accuracy may be very low. Consider a model that estimates a single parameter from the data, the sample mean. The parameter-reliability can be quite good if there are many samples. But the model-accuracy will be low if there is significant variance in the data.

One of the main challenges in building accurate models is to find a balance between error due to bias and error due to variance (known as the bias-variance tradeoff [14]). This tradeoff is intimately tied to the model complexity. Some models have a low level of complexity (few parameters). These models may *underfit* the data, because they do not have sufficient flexibility to capture the variation in the diffusion signal with the direction of measurement (Figure 1). Models with high complexity (many parameters) may *overfit* the data. These models capture the variation in the diffusion signal but they also capture the variation due to noise.

To limit the effects of overfitting on our inferences, we can compare model predictions to a second data set with independent noise samples *(cross-validation)*. Specifically, we fit a model to a first data set and then measure model-accuracy in predicting a second independent measurement. In this paper we illustrate how to measure model-accuracy for diffusion-weighted imaging data used to understand human white matter. There are many different models of within-voxel diffusion, and the number of ideas continues to expand. It is impractical to evaluate model-accuracy for all models, and thus our goals here are to (a) explain the ideas, (b) apply them to two of the most widely used diffusion models, and (c) make available software and data so that other investigators can apply our model-accuracy methods easily to other models.

## Diffusion models

A model with only a few parameters (low level of complexity) is the diffusion tensor model (DTM [15]). It approximates the data as a 3-dimensional Gaussian diffusion process. The model continues to be used in tractography algorithms [16], in diagnosing clinical conditions [17], and in characterizing behavioral variability [18–20]. Despite its widespread use, there have been no comparisons of the DTM fits with whole-brain diffusion data collected on a standard clinical scanner (see Table 1 for a list of other evaluations of DWI models available in the literature).

When the DTM was introduced, it was thought that the principal diffusion direction (PDD) of the tensor was a useful estimate of the unique orientation of fascicles, within each voxel. In fact, the PDD is not a good estimate of the local fiber direction [21–24]. For example, crossing fascicles oriented in two different directions may generate a diffusion signal whose principal diffusion direction is intermediate to the two directions, agreeing with neither fiber [25,26]. This emphasizes the importance of parameter-validity: even if the model fits the data well, researchers need to take care when interpreting the parameters of the model.

More modern models of the data increase the complexity to improve parameter-validity. They do so by adding additional parameters and going beyond the Gaussian assumption of the DTM. These additional parameters can have a variety of interpretations, and in a subset of these models, investigators interpret them as an explicit model that allows multiple fascicle orientations in a single voxel [1,3,22,23,27,28] These models contain very large numbers of model parameters. Each voxel is modeled as the sum of an isotropic signal and the weighted sum of signals from a set of fascicles at different orientations.

There is much in common among a subset of this new generation of models. First, they make explicit estimates of the number and volume of the fascicles in various directions from the DWI data. Second, they all use some means to control for the noise due to variance, and over-fitting, by means of regularization [14]. The main methods all limit the number of fascicles in the estimated solution, and for this reason we refer to this new generation of models as sparse fascicle models (SFM).

The DTM and SFM models have been assessed for parameter-reliability, but not model-accuracy [3,9,11,12,27–29]. That is, researchers often demonstrate how repeatable the fascicle direction estimates are for different acquisitions of the same data set, or for simulations in which a particular set of fascicle directions has been entered. But, accuracy (quantified as $R^2$, goodness-of-fit, prediction error, etc.) is not generally reported and there is no accepted methodology to evaluate the accuracy of the model fit.

We implemented a cross-validation framework for evaluating model-accuracy of the DTM and SFM diffusion models. We measured model-accuracy using diffusion data obtained with several b-values and high angular resolution, (Figure 1, top row). We compared model-accuracy to the repeatability of the measurements in a replication (test-retest reliability). As an example, we show the DTM fit to the diffusion data in Figure 1 (bottom row). Clearly, the DTM does not fit all of the details in the measurements. The question we ask is whether these details in the data are reliable and should be fit, or whether they should be treated as noise.

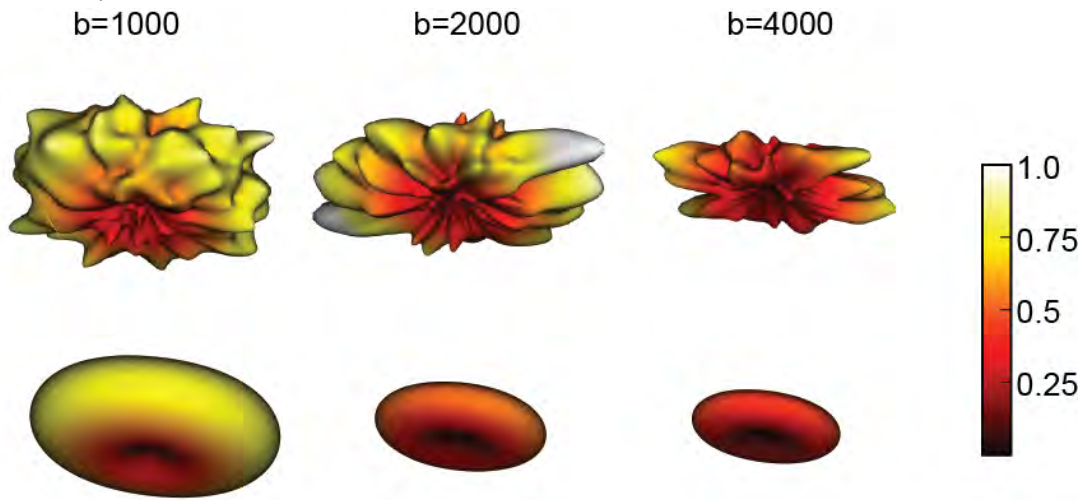

*Figure 1: The diffusion-weighted signal attenuation measured in a voxel in the corpus callosum.* *The columns show data obtained at three different b-values. (Top row) Diffusion data: The signal is interpolated on the sampling sphere. Note the differences in the spatial distribution of the signal on the sphere between the measurements obtained using different b-values. (Bottom row) DTM diffusion signal predictions: A tensor model is fit separately to the data at each b-value. The surface shows the signal predicted by the model in each direction.*

# Methods

## Subjects

Subjects were six healthy male participants, ages 27-40 (mean: 32.6). All data are available to download (see Table 2). The Stanford University Institutional Review Board approved the experimental procedures and participants provided written informed consent.

## Diffusion-weighted MRI

MRI data were collected at Stanford's Center for Cognitive and Neurobiological Imaging on 3T GE Discovery MR750 MRI system. A 32-channel head coil was used. A twice-refocused spin echo diffusion-weighted sequence [30] was used with several different acquisition schemes. Two participants were scanned with 150 different directions of diffusion-weighting. The spatial resolution of the measurement was 2x2x2 $mm^3$. In different scans, b-values were set to 1000, 2000 and 4000 $s/mm^2$ and respectively, TE values were: 83.1/93.6/106.9 msec. Ten non-diffusion weighted images (b=0) were acquired at the beginning of each scan. Two scans were performed in each b-value in immediate succession. Data at a b-value of 2000 were collected in one session and data at a b-value of 1000 and 4000 were collected in a separate session. All 6 participants were also scanned in a sequence with 96 diffusion-weighting directions at a higher spatial resolution of 1.5 x 1.5 x 1.5 $mm^3$. In this sequence only one b-value was used: 2000 $s/mm^2$ (TE=96.8 msec). Ten b=0 images were acquired at the beginning of each scan. Two sets of images were acquired in immediate succession. To mitigate the effects of EPI spatial distortions, measurements of the B0 magnetic field were performed in the high-resolution protocol. Field maps were collected in the same slices as the DWI data using a 16 shot, gradient echo spiral trajectory pulse sequence. Two volumes were successively acquired, one with TE set to 9.091 ms and one with TE increased by 2.272 ms, and the phase difference between the volumes was used as an estimate of the magnetic field. To track slow drifts in the magnetic field (e.g., due to gradient heating) field maps were collected before and after the DWI scans and between successive DWI scans. See Table 2 for a summary of the data.

## Pre-processing

MR images were motion corrected to the average b=0 image in each scan, using a rigid body alignment algorithm, implemented in SPM (http://www.fil.ion.ucl.ac.uk/spm/). The direction of the diffusion-gradient in each diffusion-weighted volume was corrected using the rotation parameters from the motion correction procedure. Because of the relatively long duration between the RF excitation and image acquisition in the twice-refocused spin echo sequence used, there is sufficient time for eddy currents to subside. Hence, eddy current correction was not applied. All pre-processing steps have been implemented in Matlab as part of the mrVista software distribution [31] which can be downloaded at http://github.com/vistalab/vistasoft. In the high-resolution protocol (1.5 mm isotropic), field maps were smoothed in space and time using local linear regression and these smoothed maps were used to unwarp the diffusion-weighted volumes, correcting for spatial distortions due to drifts in the main (B0) field [32].

## Anatomical MRI and tissue type segmentation

Segmentation of different types of tissue was performed on high-resolution T1-weighted image. Two FSPGR images were acquired at 0.7x0.7x0.7 $mm^3$ resolution and averaged to increase SNR of tissue contrast. An initial segmentation was performed using Freesurfer [33] and additional manual editing of the segmentation was then performed using itkgray [34]. The white-matter mask image was resampled to the DWI data resolution. To prevent systematic bias in voxels that are classified as white matter but contain partial volumes of CSF or of GM, we excluded voxels that have a mean diffusivity larger than 2 inter-quartiles from the median of the mean diffusivity distribution across the entire T1-defined white matter mask [35]. This segmentation process was based on the data in the b=1000 measurement for the 2x2x2 mm$^3$ protocol.

## The diffusion signal

To quantitatively describe the diffusion signal, we use the classical formulation proposed by Stejskal and Tanner [36]: the signal measured in every voxel in the brain in a spin echo experiment, may depend on the application of a second (diffusion) gradient. Suppose that the non-diffusion weighted signal in a voxel is $S_0$ and the signal measured with the application of a diffusion weighting gradient in the direction θ is $S_b(θ)$. The strength of the applied diffusion gradient, the duration of these gradients and the time interval between them are experimenter-controlled variables and will all affect the sensitivity of the measurement to diffusion in the measured volume. These are all summarized in one number: *b* [37]. The decline in signal with diffusion weighting, which results in the relative diffusion-weighted signal, $S(θ, b) = S_b(θ)/S_0$, is well-described by a decaying exponential function [36]:

$$S(θ, b) = e^{-bA(θ)} \quad (1)$$

Where $S(θ, b)$ is the relative signal measured when the diffusion-sensitizing gradients are applied in the direction θ with the parameters (magnitude, duration, etc.) b (see Figure 1, top row). The apparent diffusion coefficient, A(θ), is a direction-dependent quantity that depends on the hindrance of the diffusion of water in the direction of the applied gradient by elements of the tissue, such as cell membranes.

## The diffusion tensor model (DTM)

The diffusion tensor model (DTM; Basser et al., 1994) predicts the apparent diffusion coefficient in every direction as:

$$A(θ) = θ^t Q\, θ \quad (2)$$

where θ is a unit vector in the direction of the applied diffusion gradient and Q is a positive-definite quadratic form.

To fit the DTM we compared ordinary least-squares and weighted least-squares fitting [11], both conducted on $\log(S(θ, b))$ and non-linear least-squares fitting [38], conducted on $S(θ, b)$. All these fit methods are implemented in the freely-available dipy software library (http://nipy.org/dipy [39]). For these data, the three methods produced very similar parameter estimates and cross-validation goodness-of-fit (see Model Evaluation). Below, we present model fits obtained with the weighted least-squares method.

## The sparse fascicle model (SFM)

The family of sparse fascicle models (SFM) follow the principles first proposed by Frank [22,23]. These models have since evolved in the work of Behrens et al. [1], Dell'Acqua et al. [3] and Tournier et al. [28,27]. These models treat each MRI voxel as comprising two types of compartments: (a) non-oriented tissue that gives rise to an isotropic diffusion signal that is constant across measurement directions, and (b) a set of oriented fascicles of various volume fractions, with each fascicle giving rise to an anisotropic diffusion signal. The diffusion signal is modeled as the sum of the signals from these compartments [1]:

$$S(\theta, b) = \beta_o e^{-bD} \sum_{i=1}^{F} \beta_i e^{-b\theta^t Q_i \theta} \quad (3)$$

The weight $\beta_0$ represents the fraction of the voxel that is occupied by isotropic components with an apparent diffusivity, D. This term depends on many factors (such as the partial volume of cerebrospinal fluid in the voxel, the distribution of sizes of cellular components in the voxel, etc.). The weights $\beta_i$ are weights on individual putative fascicles in the voxel (where **F** is the number of these compartments). Each of the anisotropic components in all voxels is assumed to be well-represented by the response function of a canonical tensor that is used as a kernel function in all of the white matter [3,27,40–42]. In the work of Behrens et al. [1], the kernel has a specific form (axial diffusivity equal to 1 and radial diffusivities both equal to 0). Another approach, which we adopt, is to estimate the form of the kernel from the data [27]. Specifically, we estimated a kernel for each scan and each subject from a region of interest (ROI) in the corpus callosum (CC). The CC was chosen because it contains axons oriented in a single direction, and thus approximates the kernel function. The ROI was defined using an automated method, which selects voxels based on the direction of the principal diffusion direction (left-right), high FA (>0.4), a target MD range (between 0.7 mm$^2$/s and 1.1 mm$^2$/s) and uniformity of the b=0 signal across the population of voxels in the estimated position of the CC [39,43]. The 250 most linear voxels in the CC ROI were identified and the median axial and radial diffusivity in this collection was chosen to represent the axial and radial diffusivities in the canonical tensor.

Equation 3 can be rewritten to better distinguish the isotropic and anisotropic components of the signal. First, we calculate the direction-dependent deviation of the signal, removing the mean of the fiber response function modeled by the canonical tensor.

$$O_i(\theta, b) = e^{-b\theta^t Q_i \theta} - \mu_i$$

$$\text{where:} \quad \mu_i = \frac{1}{T}\sum_{j=1}^{T} \beta_i e^{-b\theta_j^t Q_i \theta_j} \quad (4)$$

The $O_i$ term is the fiber orientation modulation (fOM). This is the directionally-dependent signal from a single estimated fascicle with tensor $\mathbf{Q_i}$. The tensor $\mathbf{Q_i}$ is a rotated version of the canonical tensor and models a single coherent population of white matter fibers. The quantity $\mu_i$ approximates the mean signal (across measurement directions, j = 1…T), which is independent of i. The computed value of $\mu_i$ varies by <0.5% between different i, due to the discrete sampling of the sphere. The variation is very small, because the sampling density is sufficiently high. Hence we drop the subscript on $\mu$ in the following.

Using simple algebraic manipulations, we can rewrite Equation 3 into the sum of one anisotropic term and a set of fOM terms:

$$S(\theta, b) = \beta_o e^{-bD} \sum_{i=1}^{F}(\beta_i e^{-b\theta^t Q_i \theta} + \mu - \mu) \qquad (5)$$

$$= \left[\beta_o e^{-bD} + \mu \sum_{i=1}^{F} \beta_i\right] + \sum_{i=1}^{F} \beta_i O_i$$

$$= W_0 + \sum_{i=1}^{F} \beta_i O_i$$

For any fixed b value, the term $W_0$ in equation 5 is constant, independent of direction. The fOM terms, $O_i(\theta)$, are zero-mean. Thus, $W_0$ is equal to the mean of $S(\theta)$ across directions, which we denote $\bar{S}$. Re-expressing the isotropic component as the mean signal and rearranging terms, we can rewrite Equation 3 as:

$$S(\theta) - \bar{S} = \sum_{i=1}^{F} \beta_i O_i(\theta) \qquad (6)$$

We estimate the weights, $\beta_i$ from the diffusion signal in a voxel using Equation (6).

## Matrix representation of the SFM

In this section we introduce and solve a matrix representation of Equation 6. Every voxel contains an isotropic component and a set of anisotropic components. As shown above in Equations 5 and 6, the isotropic component is equal to the mean of the signal. The contributions of **F** anisotropic components are fit solving a linear regression problem, which we express in matrix form:

$$(7) \quad \begin{pmatrix} S(\theta_1) - \bar{S} \\ S(\theta_2) - \bar{S} \\ \vdots \\ S(\theta_T) - \bar{S} \end{pmatrix} = \begin{pmatrix} O_1(\theta_1) & O_2(\theta_1) & \cdots & O_F(\theta_1) \\ O_1(\theta_2) & O_2(\theta_2) & & O_F(\theta_2) \\ \vdots & & \ddots & \vdots \\ O_1(\theta_T) & O_2(\theta_T) & \cdots & O_F(\theta_T) \end{pmatrix} \begin{pmatrix} \beta_1 \\ \beta_2 \\ \vdots \\ \beta_F \end{pmatrix}$$

9Where the columns $O_i(\theta_j)$ denotes the fOM of the $i^{th}$ fascicle (i = 1…F), in the $j^{th}$ measurement direction ($\theta_j$, j = 1…T). We express the Equation (7) concisely as $\mathbf{s} = \mathbf{X}\boldsymbol{\beta}$, where $\boldsymbol{\beta}$ is the weight vector (F elements), $\mathbf{s}$ is the mean-removed relative signal vector (T elements) and $\mathbf{X}$ is the regression matrix (T x F elements).

## Solving the SFM equations

We would like a sparse solution to Equation (7). Specifically, we would like to choose the minimal number of fascicles that best represent the data observed. We obtain this sparse solution and control for over-fitting using Elastic Net [44]. By requiring a sparse solution, it is straightforward to solve the under-determined equation in which the number of columns (fascicles) exceeds the number of rows (measurement directions; F > T).

The Elastic Net algorithm solves $\mathbf{s} = \mathbf{X}\boldsymbol{\beta}$ for $\boldsymbol{\beta}$, while minimizing the following penalty:

$$\sum_{i=1}^{T}(S_i - \hat{S}_i)^2 + \lambda \sum_{j=1}^{F}(\alpha \beta_j^2 + (1-\alpha)|\beta_j|) \qquad (8)$$

The first term is the sum of squared error between the measured signal, $S$, and the model-estimated signal, $\hat{S}$; the second term contains two regularization components (Elastic Net penalty; (Zou and Hastie, 2005)). The first component penalizes for the sum of the squares of the weights (L2 norm) and the second penalizes for the sum of the absolute value of the weights (L1 norm). The scalars $\boldsymbol{\lambda}$ and $\boldsymbol{\alpha}$ are regularization parameters. Setting a high value of $\boldsymbol{\lambda}$ induces a solution that conforms more to the regularization constraints at the cost of reducing the fit to the data. The parameter $\boldsymbol{\alpha}$ varies between 0 and 1: when $\boldsymbol{\alpha}$ = 0, the algorithm provides a solution equivalent to that provided by the Lasso algorithm [45] and penalizes the weights in β by an L1-norm. When $\boldsymbol{\alpha}$ =1 the algorithm is equivalent to ridge regression (also known as Tikhonov regularization; [46]) and penalizes the weights in β by an L2-norm. Values of $\boldsymbol{\alpha}$ between these two values emphasize one or the other. We chose the values of $\boldsymbol{\lambda}$ and $\boldsymbol{\alpha}$ that provided the smallest median cross-validated error (across voxels in the white matter) in predicting the diffusion data (see S1 Figure).

## Model accuracy

To estimate model-accuracy, we compute the goodness of fit between the prediction of a model and the measurement. To estimate the difference between two sets of measurements, or between a measurement and a model prediction, we use a root mean square error (RMSE) metric

$$RMSE(x,y) = \sqrt{\frac{\sum_{i=1}^{T}(x_i-y_i)^2}{T}} \qquad (9)$$

Where **x** and **y** are two different measurements in a white matter voxel or a measurement and a model prediction in that voxel over all the diffusion weighted directions **θ**$_i$ (T=150 directions in our measurements).

To estimate the goodness-of-fit of a model we use cross-validation. Specifically, we estimate the parameters of the model on one set of data; we then use the model parameters to predict the signal in a second, independent data set. We use RMSE of the signal as a measure of model accuracy. We further assess model-accuracy by comparing the model RMSE to the RMSE of repeated measurements: test-retest reliability. DWI data was collected twice in each b-value and test-retest reliability is calculated as the RMSE between the two measurements in each voxel across directions of measurement. RMSE is given in the units of the measurement. In the case of MRI data, these are the scanner signal units, which do not have a straightforward physical interpretation. Thus, it is difficult to compare RMSE values across different locations in the brain and across different measurement parameters. The distribution of RMSE of test-retest reliability is very similar across b-values (see Results, Figure 2). We speculate that this indicates that the noise arises principally from sources that are independent of the diffusion itself, including subject motion, thermal changes in the scanner equipment, and physiological noise. Nevertheless, RMSE does not provide a natural benchmark. Less error is better, but it is not clear how small of an error is good enough.

To create a meaningful measure, we normalize the RMSE of the model prediction on a second data set to test-retest reliability. That is, we normalize to the RMS of the difference between the two measurements (test-retest reliability). We compute a measure of relative *RMSE* (*rRMSE*):

$$\text{rRMSE} = \frac{(\text{RMSE}(M1,D2)+\text{RMSE}(M2,D1))}{2\text{RMSE}(D1,D2)} \qquad (10)$$

where **D1** is the diffusion-weighted signals measured in the first data set and **M1** are the signals predicted from the model fit to this data, and similarly for the second data set, **D2**. The measure provides us with an index of the goodness-of-fit of a particular model, relative to the reliability of the measurement.

The denominator in Equation 10 is an indication of test-retest reliability. If a model is exactly as accurate as test-retest reliability, the numerator is identical to the denominator, and rRMSE has an expected value of 1. This is the goodness-of-fit of the null model that the data will repeat itself exactly. If a model has higher model-accuracy than test-retest reliability, RMSE is smaller than the denominator in Equation 10, and rRMSE is less than one. This means that the model predicts the replication measurement more accurately than the original data would (more accurate than test-retest reliability). Hence, this measure provides a natural quality scale for models fit to different data: when the value of rRMSE is smaller than 1, the model is more accurate than test-retest reliability. For the simple case of IID signals, with zero-mean Gaussian noise and standard deviation, σ, if the model M perfectly predicts the data, the rRMSE has an expected value of $\frac{std(M-D1)}{std(D1-D2)} = \frac{\sigma}{\sigma\sqrt{2}} = \frac{1}{\sqrt{2}}$. Hence, a perfect model has an expected value of the rRMSE of 0.707.

Due to subject motion between volume acquisitions, the diffusion gradient for that measurement is rotated slightly from the direction programmed into the sequence; we account for the rotation of this vector by a motion correction procedure that aligns each diffusion scan to the mean non diffusion-weighted scan [47]. We use the motion-corrected gradient directions in each scan to predict the signal in that scan. To calculate RMSE(D1,D2), we treated the two sets of gradient directions as equal, so that deviations from this assumption are part of the measurement noise. As a practical matter, for these subjects and conditions the differences in direction gradients are very small (maximum deviation less than 2 deg).

## Simulations of fiber crossings

To further explore the differences between model-accuracy, model reliability, and parameter-validity, and the distinction between model fitting and model interpretation, we evaluated the model-accuracy and parameter-validity of the two models (DTM and SFM) in synthetic data generated through numerical simulations of different tissue configurations. In each voxel, we simulated a signal assuming that each fascicle in the voxel would generate a signal that can be approximately described by a tensor with a principal diffusion direction aligned along the direction of the fascicle. The diffusion signal was then generated as a weighted linear combination of fascicle signals from two fascicles in each voxel. The relative orientation of the fascicles was varied, such that the crossing angles between the fascicles were between 0 and 90 degrees and the relative contribution of the fascicles varied from a 1:1 ratio to 1:0 (single fascicle).

Noise was added to each voxel based on the actual noise in the measurements. We computed the noise in each voxel in the white matter as:

$$\text{Noise}(\theta) = \frac{D1(\theta) - D2(\theta)}{2} \qquad (11)$$

Where $D1(\theta), D2(\theta)$ are the diffusion signals in each direction in the two different measurements. In each iteration of the simulation, we randomly chose the noise from one of the voxels in the white matter and added this noise sample to the simulated signal. We rectified the signal to be non-negative after addition of the noise.

We fit the DTM and the SFM to the signal in each simulation. To assess parameter-validity, we computed how well they represented the original tissue configuration entered into the simulation. For the DTM, this was done by calculating the minimal angular difference between the DTM principal diffusion direction (PDD) and one of the simulated fascicle directions. For the SFM, the minimal angular difference was calculated between each direction for which there was a non-zero weight and the median of these minimal angular differences was computed. The angular difference between the PDD (defined as the direction of the largest eigenvector for the DTM, and the direction of the largest parameter for the SFM) in two iterations of the simulation was computed. For each crossing angle and each ratio of fascicle contributions, we performed 500 simulations and computed both model-accuracy and parameter-validity for b=1000, 2000 and 4000 s/mm2.

### Estimating the effects of number of measurements

One important application of an estimate of model-accuracy is to select among competing models, such as the DTM and the SFM in a particular data set. Another application of this estimate is the evaluation and comparison of different measurement schemes. To estimate the effects of different measurement schemes on model-accuracy, we sub-sampled the 150 directions of measurements in the data, to different numbers of measurements. To guarantee that the measurements in each number *n* of sub-samples were in maximally separated directions [10], we first chose one of the 150 measurement directions to be the origin. The set of electro-static repulsion points with *n* vectors was then aligned by rotation to this vector. For each of the subsequent n-1 points in this electro-static repulsion point set, the experimental point that was closest (smallest angle) was added to the set. In each iteration, the selected vector was then removed from the candidate pool, to avoid repetition, before continuing to the next iteration of the selection process. The process ended when n vectors were chosen. Relative RMSE was calculated for this sub-sample of measurement points in every voxel in the white matter.

Reproducible research

To facilitate the reproducibility of these results [48] we provide a full implementation of the analysis that led to each of the figures and all of the conclusions in the text at http://github.com/vistalab/osmosis/).

# Results

### Signal and noise in DWI measurements

Noise in the DWI measurements inherently limits the fit of models to the data. To assess the noise in the DWI measurements, we calculated test-retest reliability, as the root of the mean squared error (RMSE) between two measurements, RMSE(D1,D2). The distribution of RMSE(D1, D2) across all of the voxels in the white matter does not differ substantially between measurements conducted at different b-values (Figure 2), suggesting that the noise mainly arises from sources that are common across the measurements in different b-values (subject motion, physiological noise, etc.).

We also estimated the SNR in each voxel, $\text{SNR} = \frac{\mu}{\sigma}$, where μ is the average of the signal across the diffusion-weighted measurements and σ is estimated from the standard deviation in the non diffusion-weighted measurements (10 for each b-value measurement). We correct the computed value of σ for bias due to the small sample size (For a detailed proof of this correction see http://nbviewer.ipython.org/4287207). The SNR decreases as b-value increases, because the signal decreases with b-value.

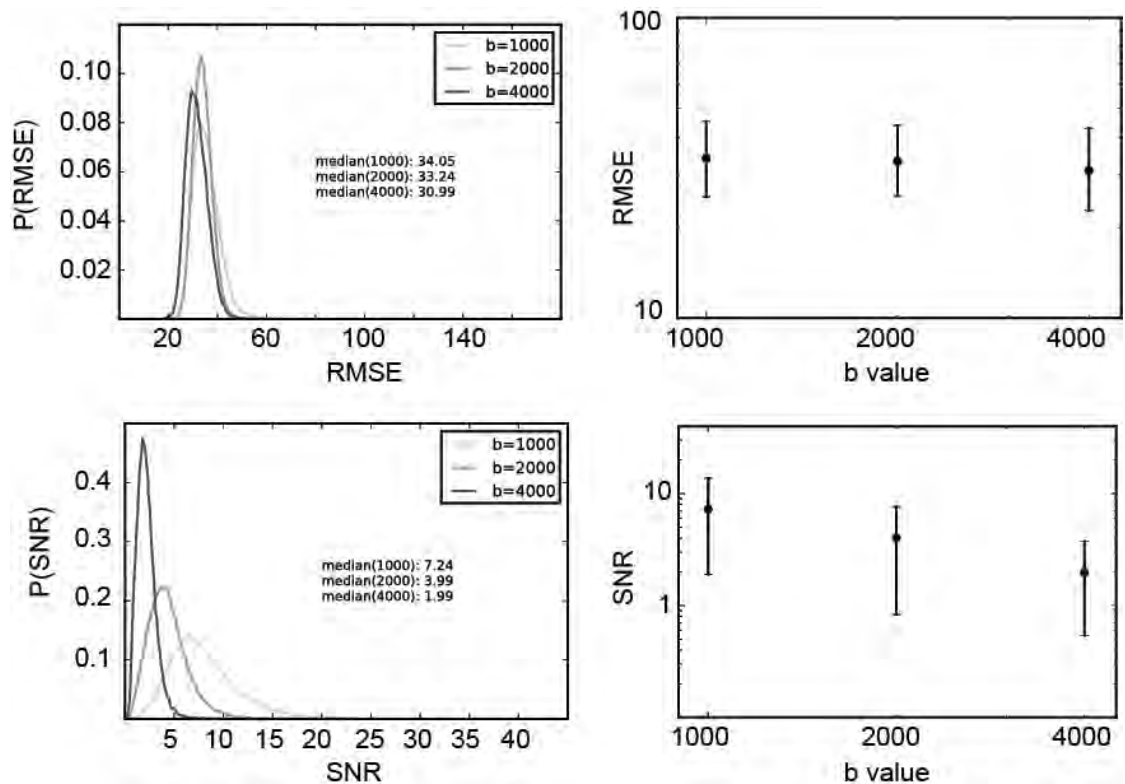

*Figure 2: RMSE and SNR of diffusion MRI measurements.* Error bars delineate the 95% interquantile range. RMSE does not change across b-values, but SNR changes substantially, with the median decreasing from approximately 7 (b =1000 s/mm$^2$) to approximately 2 (b=2000 s/mm$^2$).

## DTM cross-validated model-accuracy is better than test-retest reliability

We fit each model to one data set and evaluated how well the model predicts a second, independent data set. The quality of the prediction is shown in several ways (Figure 3). The two scatter plots analyze the data from a typical voxel in the corpus callosum in one individual, at a b-value of 2000 s/mm$^2$. Panel A shows the repeatability of the measurements in this voxel, RMSE(D1,D2). Panel B shows the prediction used to calculate RMSE(M2,D1); this scatter plot is very similar to the symmetric prediction, RMSE(M1,D2).

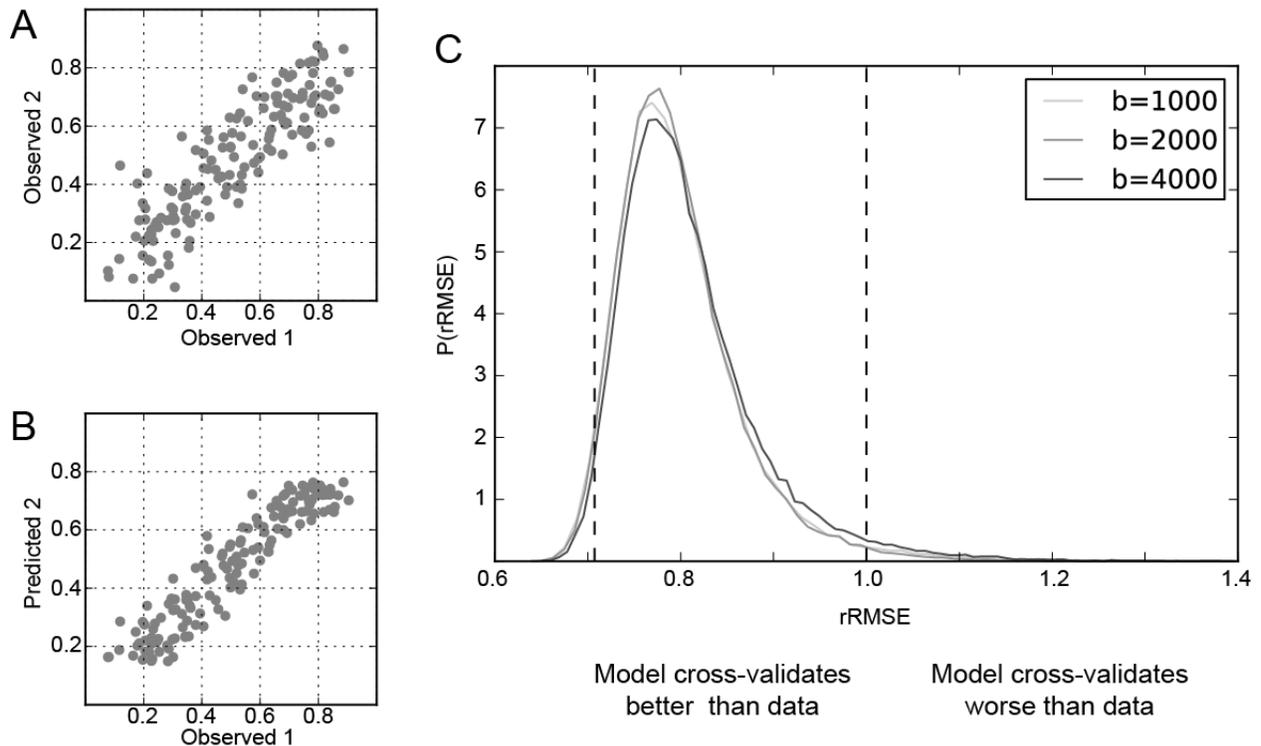

*Figure 3: The diffusion tensor model cross-validates to an independent data set better than the data cross-validate. (A) The relative diffusion-attenuated signals $S(\theta, b)$ in a single voxel in two measurements are compared. Each point in the scatter-plot represents the repeated measurement in one of 150 diffusion directions. (B) The signal measured in the one data set is compared to the predicted signal from fitting a tensor model to the other data set. (C) The distribution of rRMSE values in the white matter for the diffusion tensor model (DTM). The rRMSE is calculated for each voxel as ratio of the RMSE in (B) (model prediction vs. data) divided by the RMSE in (A) (test-retest reliability). When values of rRMSE are smaller than 1 (right dashed line), the DTM better predicts a subsequent data set than repeated measurement. An optimal model will have an rRMSE distribution centered on $\frac{1}{\sqrt{2}}$ (left dashed line). Different curves show measurements at different b-values.*

Figure 3C shows model-accuracy for all of the white matter voxels and all three b-values in this same individual as histograms of voxel rRMSE values. The DTM predicts the measurements in most white matter voxels and all b-values better than test-retest reliability; that is, a large majority of the three rRMSE distributions is less than 1. For a b-value of 1000, 98%, of white matter voxels have an rRMSE smaller than 1 (median 0.78), for a b-value of 2000, 99% (median 0.78) and for a b-value of 4000, 98% (median 0.79). The 95% confidence interval on the median estimates is 0.001 (estimated by bootstrap). The median rRMSE is close to the expected value of a perfect model ($\frac{1}{\sqrt{2}}$) with only small room for improvement. An essentially identical pattern of results was observed in the second participant for whom measurements were conducted in these three b-values. Similarly, in measurements conducted in 6 participants (including these two participants) with a different spatial resolution (1.5 mm isotropic), and with a different number of measurement directions (96 directions) at a b-value of 2000 s/mm$^2$, we found that the median rRMSE for DTM was less than 1.0 in more than 99% of the voxels for all participants, and the average (across subjects) of the median rRMSE (across the white matter) was 0.78 (+/- 0.014, SD)

## The DTM model-accuracy is lowest in specific white matter regions

Next, we examine the parts of the brain in which the DTM model-accuracy compared to test-retest reliability is lowest. Voxels in which DTM rRMSE is higher than 1 are located in two major clusters (Figure 4). The rRMSE of these voxels increases with higher b-value. One of these regions is a part of the brain known to contain fascicle crossings: the centrum semiovale at the intersection of the cortico-spinal tract, running in the superior-inferior direction, the superior longitudinal fasciculus (SLF), running in the anterior-posterior direction and fascicles from the corpus callosum, running in the medial-lateral direction. In addition, with higher b-values, the model increasingly fails to account for the diffusion data in parts of the brain surrounding the optic radiations (OR).

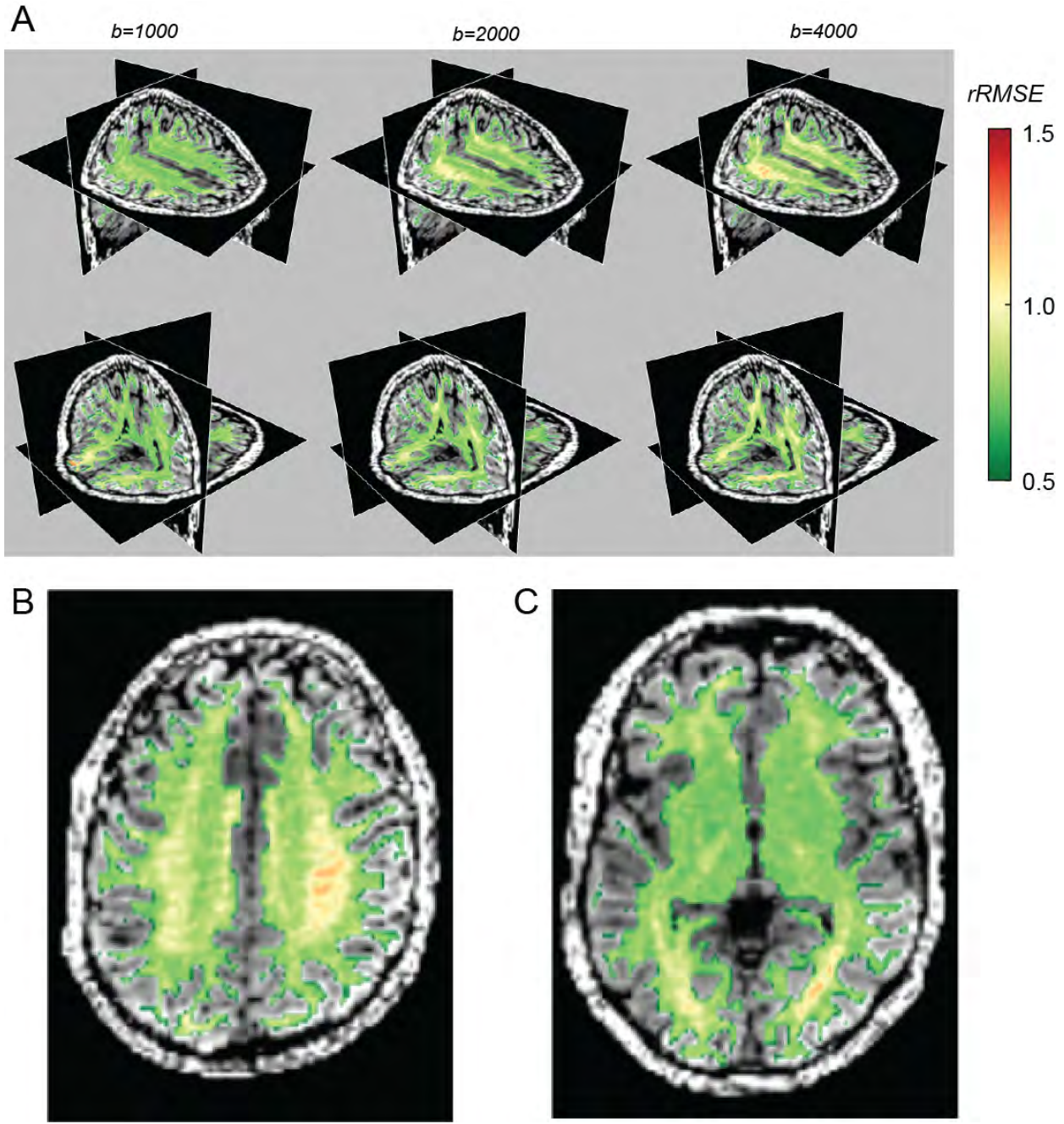

*Figure 4:* DTM model-accuracy is high (low rRMSE) in a large portion of the white matter, but systematically deviates in two locations (3 % of the voxels): the centrum semiovale (A, top row) and the optic radiation (A, bottom row). The columns show data obtained at three different b-values (b=1000, 2000, 4000) in one individual, and this pattern is observed in a second individual as well. The color overlay measures the rRMSE. Poor cross-validation (rRMSE > 1) is denoted by the yellow-red colors. (B, C) The color overlays are rRMSE maps calculated at b=4000. The two images illustrate the poor fits in the optic radiation (B) and the centrum semiovale (C).

The complexity of the diffusion signal in these locations was described in previous work. Alexander et al. [25] found that these regions were more accurately modeled with a higher order of spherical harmonic basis functions. Using nested model comparison (ANOVA) rather than cross-validation, they estimate that 5% of the voxels in the brain require an order 4 or higher spherical harmonic basis set to represent the signal (at b=1000 s/mm$^2$), noting in particular the OR.

## SFM model-accuracy

The SFM fits the data very accurately (Figure 5). The r*RMSE* is smaller than 1.0 in in 98.1% (b=1000, median 0.77), 99.9% (b=2000, median 0.76) and 99.9% (b=4000, median 0.76) of the white matter voxels in one individual, and essentially identical results were obtained in a second individual. The 95% confidence interval on the median estimates is 0.001 (estimated by bootstrap). Similar results were obtained in 6 participants (including these two), with higher resolution measurements, at b=2000 s/mm$^2$. In all participants, more than 99.9% of the voxels in the white matter have an rRMSE smaller than 1.0, and the average of the median rRMSE across white matter voxels is: 0.77 (+/- 0.013, SD).

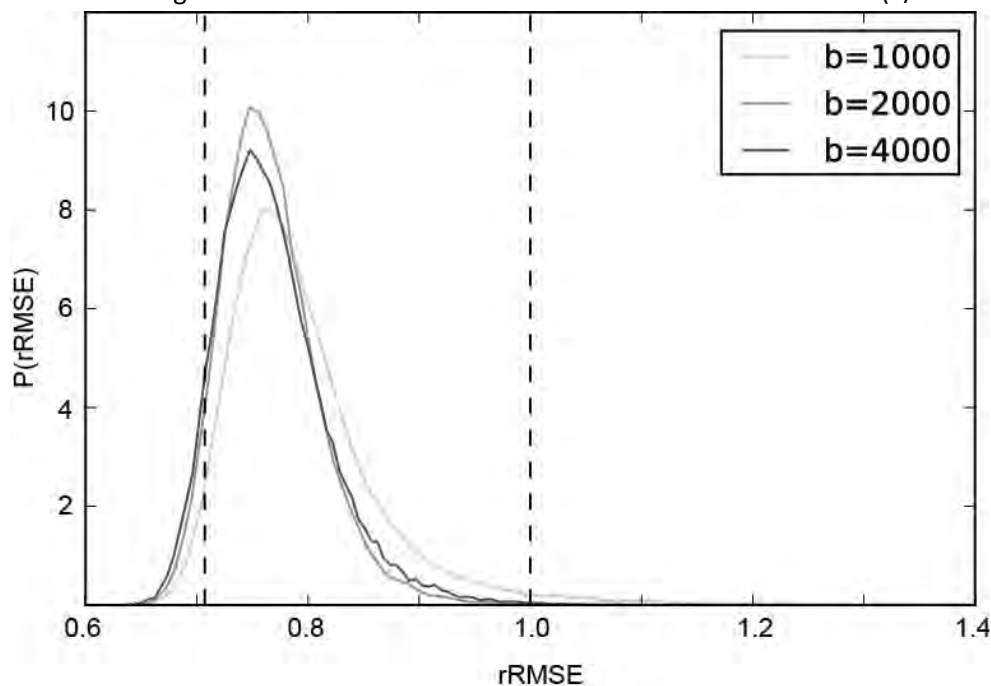

*Figure 5: The sparse fascicle model (SFM) predicts the data better than data-to-data repeatability in almost all voxels in the white matter.* For higher b-values (2000, 4000) more than 99.9% of the white matter voxels have rRMSE<1.

Through explorations of the data, we find that an excellent fit is obtained when the weight on the regularizing term is rather small (S1 Figure). The cross-validation procedure informs us that we can trust the SFM to describe the reliable features of the data.

The SFM model-accuracy is slightly better than that of the DTM almost everywhere, and in particular for b-values larger than 1000 (Figure 6). It specifically improves the quality of the fit in the regions where the DTM model-accuracy is lowest (Figure 7; compare with Figure 4). For a particular example of the differences between the SFM and the DTM, we examine the signal and the model predictions in a voxel in the centrum semiovale below.

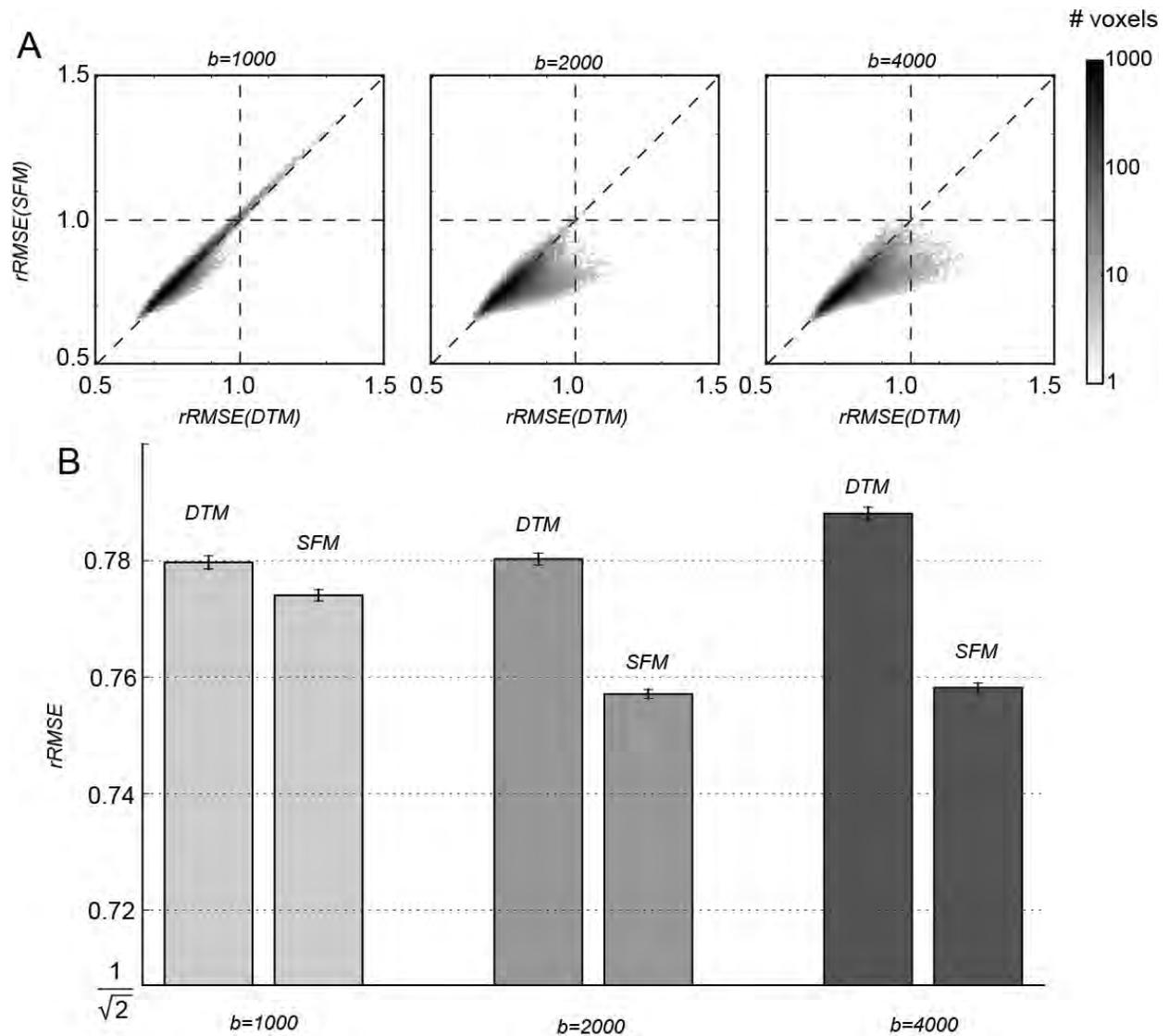

*Figure 6: The SFM fits the data better than the DTM. (A) Image histograms comparing the rRMSE of the SFM and DTM in each white matter voxel. (B) Median rRMSE of the DTM and SFM +/- 95% confidence interval estimated with a bootstrapping procedure.*

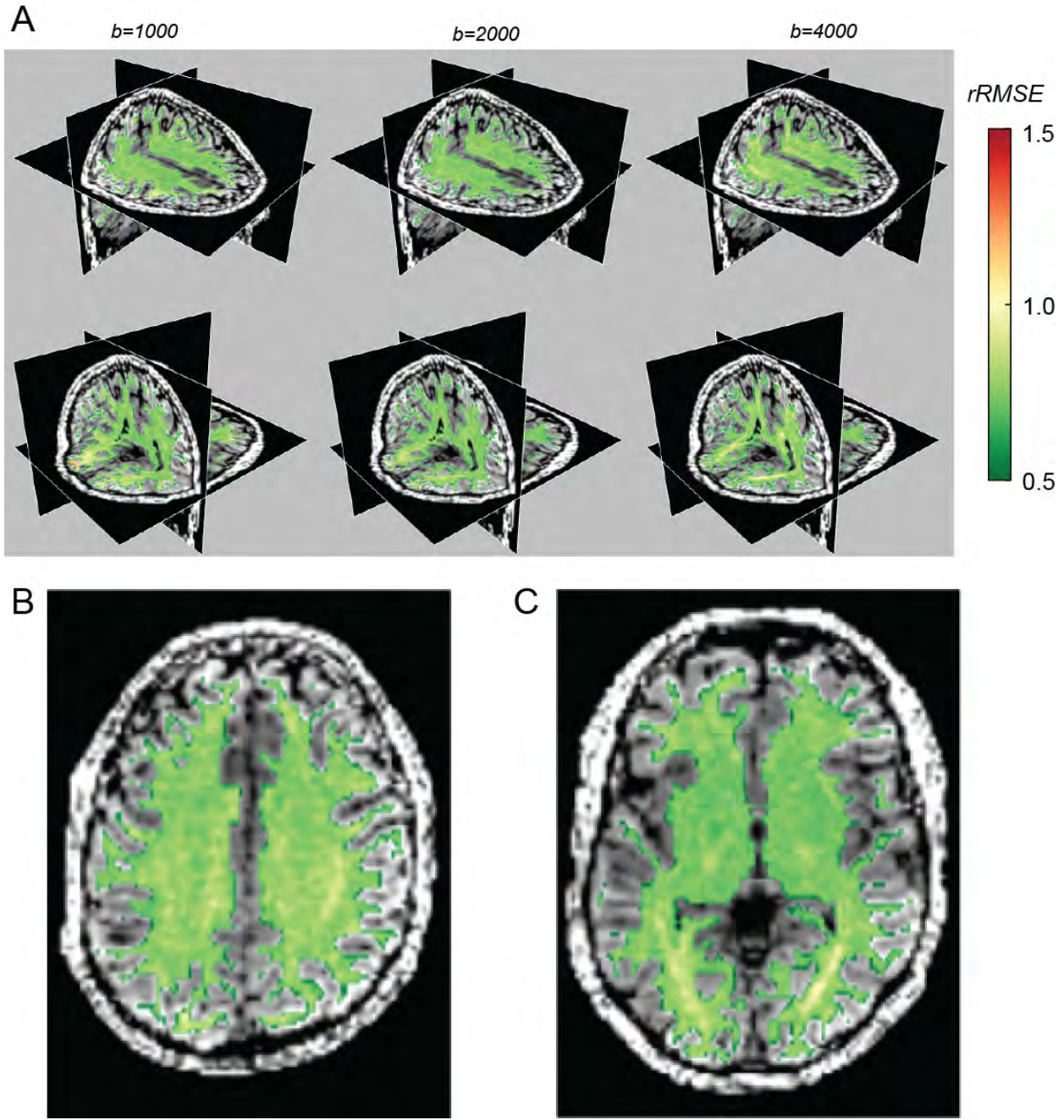

***Figure 7: The SFM substantially improves the signal prediction in the three percent of the voxels where the DTM is less reliable than the data.*** *It improves the fit in most voxels somewhat. Other details as in Figure 3. Essentially identical results were obtained in a second individual.*

## Why SFM model-accuracy is higher than that of DTM

To understand why the SFM improves the fit, we examined the signal in voxels that are substantially better fit by the SFM than the DTM. We illustrate a typical case using a voxel in centrum semiovale.

Measured with a b-value of 4000, the DTM rRMSE is 1.2 and the SFM rRMSE is 0.8. We show the interpolated diffusion signal surface in Figure 8. At low b-value (1000) the two measurements of the surface are grossly the same and the noise appears as small modulations of that surface. The signal varies slowly with angle, while the noise varies relatively rapidly with angle. At higher b-values (2000, 4000) the signal is smaller and the noise is approximately the same (Figure 1). Even so, at higher b-values the signal angular resolution is higher, and reliable features of the angular distribution emerge. For example, two reliable 'dimples' appear in the signal profiles of both measurements (Figure 8), indicating two fascicles crossing through this voxel. These 'dimples' cannot be captured by the DTM, but they are accurately captured by the SFM. Hence, the SFM outperforms the DTM for this voxel.

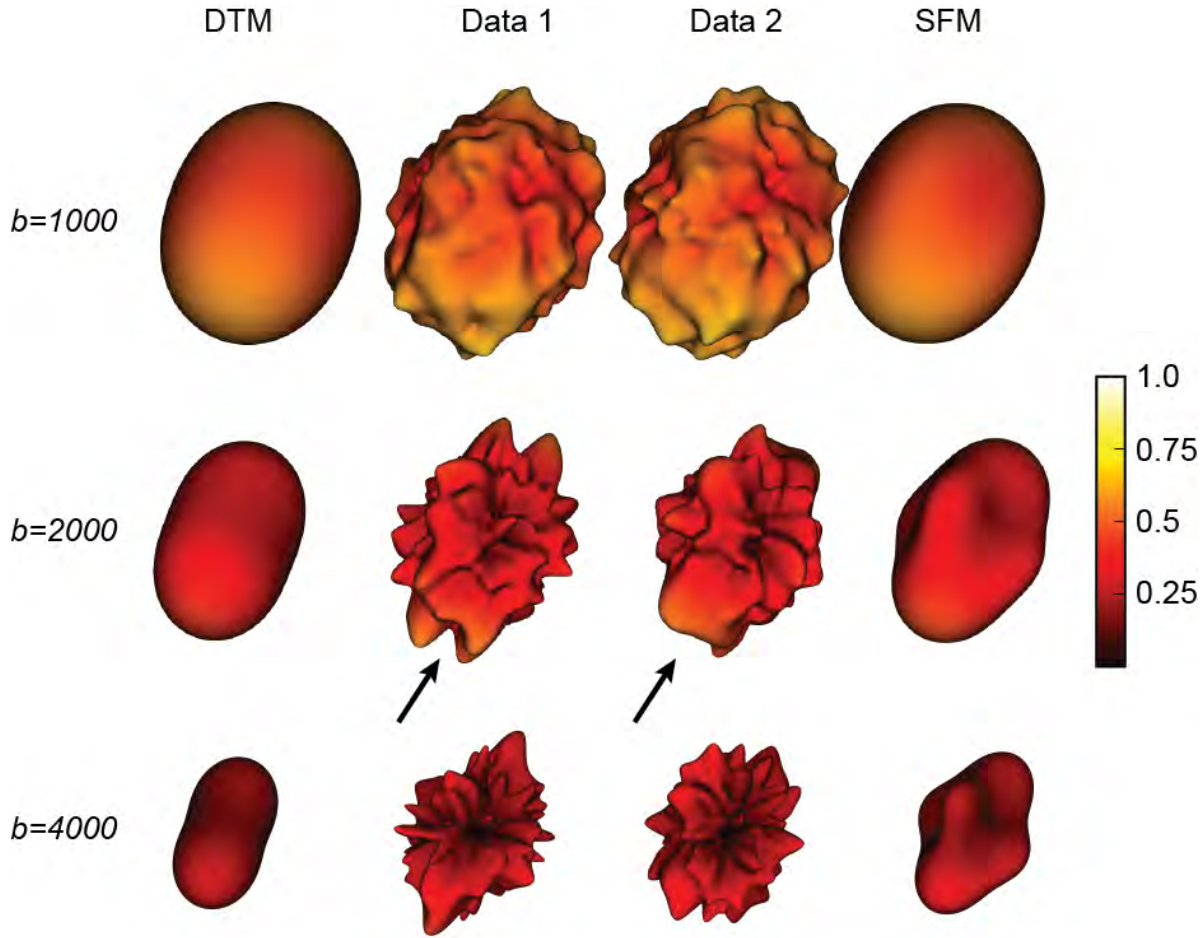

*Figure 8: Local extrema in the diffusion signal attenuation do not cross-validate well. The two middle columns are independent measurements of the same voxel from the centrum semiovale. The three rows show measurements of this voxel obtained at b = 1000, 2000, and 4000. Notice that local minima and maxima differ between replications (arrows). The DTM (left column) and SFM (right column) predictions generally cross-validate well and are much smoother than the data. This particular voxel was chosen to illustrate a case where there are likely to be crossing fascicles. At this location and at b=4000, the rRMSE of the DTM is greater than 1, while the rRMSE of the SFM is less than 1.*

The data and model fits in Figure 8 are very revealing in a second way. The SFM model predicts the independent data set better than test-retest reliability. Yet, the SFM model captures only a small subset of the features of the data (b=4000). The many small variations in the diffusion signal measured at high b-values cannot be trusted to replicate. The fitting procedure in the SFM guides the model to select those features that are reliable across data replications.

## The relationship between model-accuracy, parameter-reliability and parameter-validity

The model-accuracy differences between the DTM and the SFM are small but consistent. What are the implications of these results to the interpretation of model parameters? The estimated directions in the SFM model are supposed to correspond to the fiber orientation distribution function. The principal diffusion direction in the DTM model is not generally through to represent the fiber orientations, though in some regions of high anisotropy it is taken as the main fiber direction. Here, we quantify these ideas.

We simulated the diffusion signal in a simple case and estimated parameter-reliability and parameter-validity from these simulated signals. Specifically, we varied the angle between two fascicles passing through a simulation voxel and the relative weight of each of the two fascicles. In each voxel, realistic noise was added from the diffusion data.

Overall parameter-reliability is very good for both the DTM and SFM (Figure 9A). There is one exception: the DTM is unreliable for crossing configurations at 90 degrees. This is because in this case, for the true signal, the principal diffusion direction can be any direction along an equator and so is determined entirely by small biases induced by the particulars of the noise sample in each measurement.

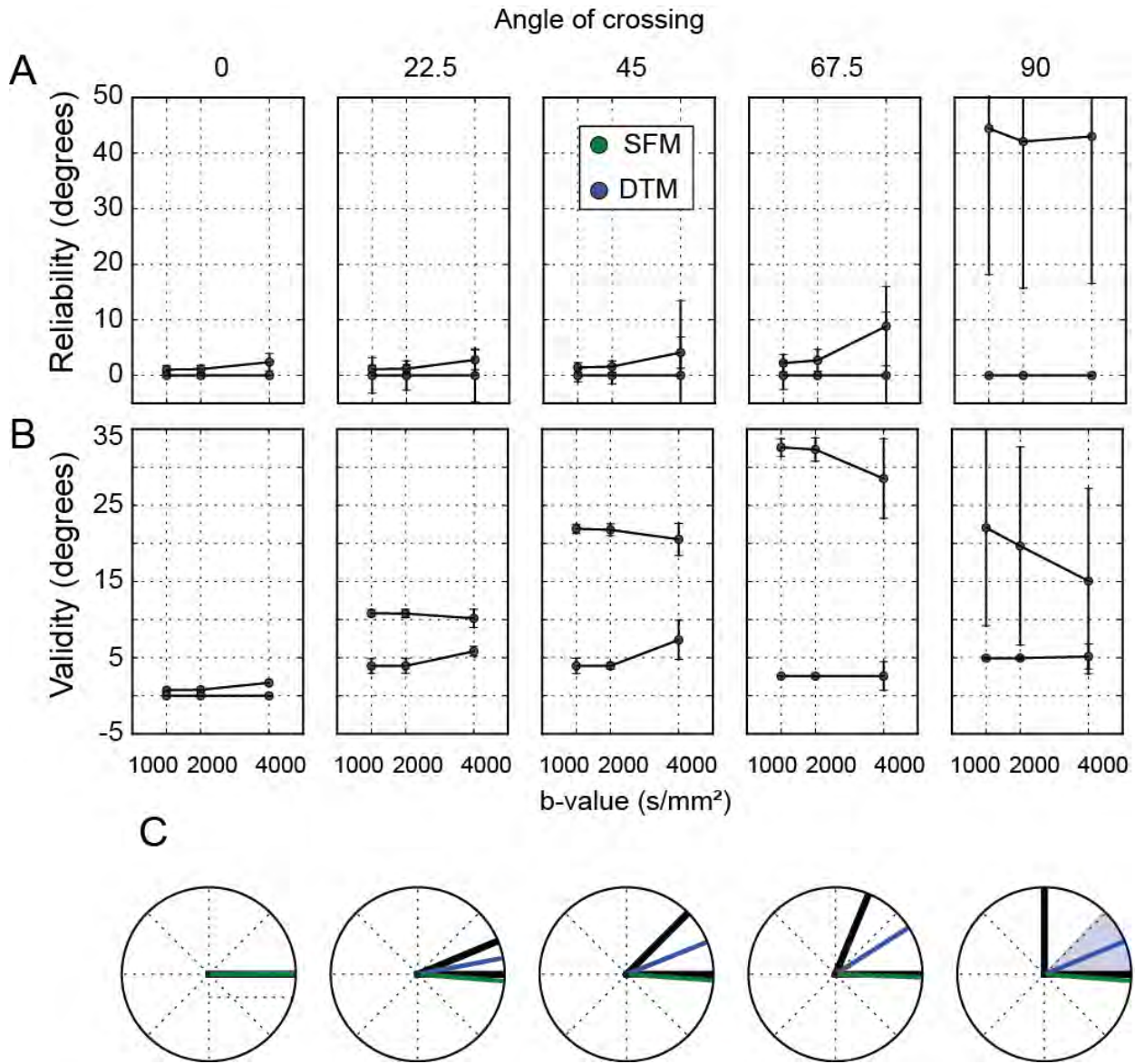

*Figure 9: A simulation study of parameter-validity and parameter-reliability of fiber ODF estimates. (A) Parameter-reliability of the DTM and SFM estimates is defined as the angular difference of the PDD between model parameters in two simulations of the same fascicle configuration with different noise. (B) Parameter-validity is estimated by examining the angular difference between the peaks of the estimated and true fODF entered in the simulation. (C) Summary of parameter-reliability and parameter-validity. The black lines represent the true simulation fascicle directions and colored lines represent the difference between the estimated and the true fODF peak (parameter-validity). The shaded region represents parameter-reliability in estimating the peak of the fODF with different noise samples. The DTM PDD is an invalid estimate of the fiber directions over a wide range of crossing angles and unreliable when crossing angles are near 90 degrees. The SFM provides a valid estimate of fiber directions, and is reliable throughout.*

Parameter-validity is assessed with regard to the directions entered in the fODF that generates the simulated signal. In this regard, the SFM performs well, while the DTM error increases with larger crossing angles (Figure 9B). This is because in cases of crossing, the DTM assigns the PDD to be an angular average of the actual directions of the fascicles entered in the simulation. The validity of the DTM model decreases with increasing crossing angle of the simulated fibers because the DTM PDD falls between the two fibers, and this is further and further away from the simulated fibers as the crossing angle grows. In a crossing angle of 90 degrees, the DTM predicts an approximately disc-like tensor, and the PDD depends on the noise, rather than on the signal. For that reason, validity is low, and the error bars are large. The validity of the SFM model does not vary substantially with the crossing angle, because it is equally able to capture the true simulated fiber directions in all crossing angles.

## Implications for experimental measurements: How many directions should we measure?

Model-accuracy is a useful measure for choosing the number of measurement directions. Previous research used simulations to demonstrate that under realistic noise conditions, DTM parameter-reliability and parameter-validity stabilizes between 20 and 30 measurements [10]. However, the previous studies did not assess the effect of this experimental choice on model-accuracy. Model-accuracy monotonically increases (rRMSE decreases) with the number of measurements for all b-values, reaching near-asymptotic levels at approximately 40 measurements (Figure 10); further increases in the number of directions provides small gains in accuracy. Considering that the DTM has only 6 independent parameters, and the SFM can have many more parameters, we may have hypothesized that the DTM would be more accurate for small numbers of directions. But in fact, the SFM median rRMSE is equal or lower than that of the DTM even when few directions are included.

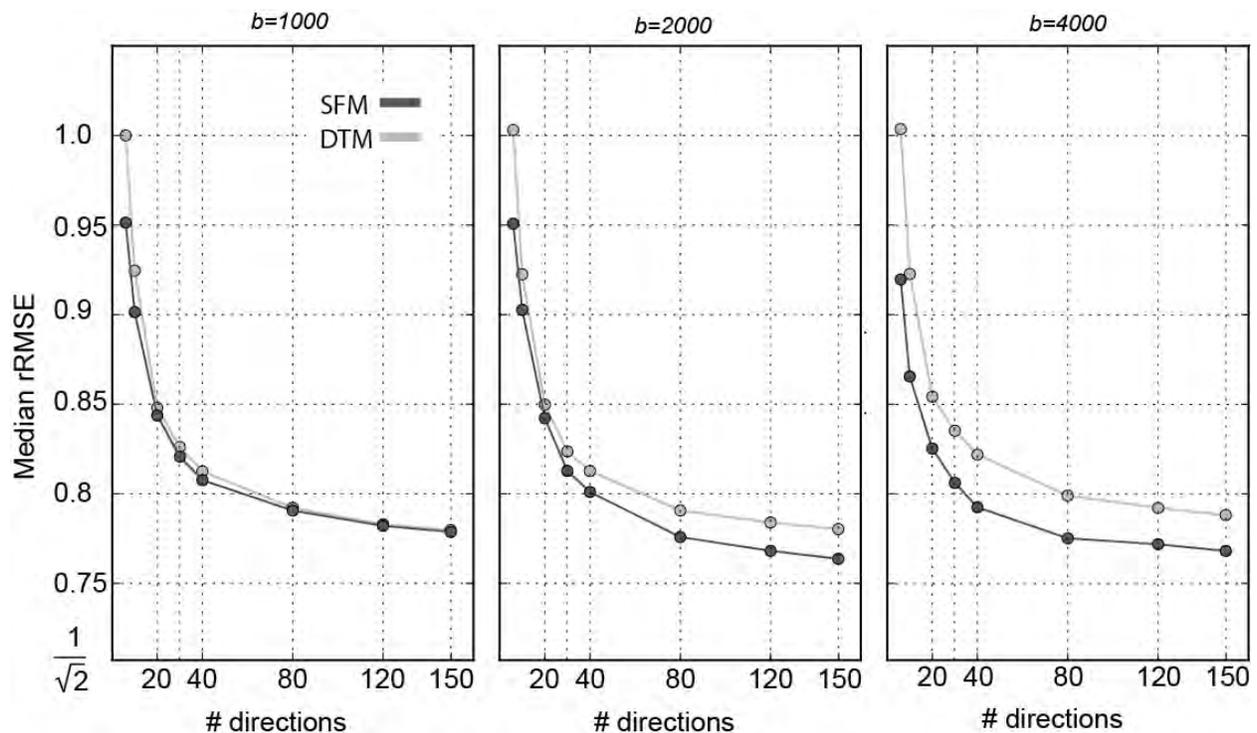

*Figure 10: The effects of number of measurement directions on model-accuracy.* Subsamples from the 150 direction measurements were used to fit the DTM (bright grey) and SFM (dark grey) and to estimate rRMSE. Median in one participant is presented. Similar results are found in a second participant, and in 6 participants in b-value of 2000 s/mm$^2$.

# Discussion

We quantitatively compared two major classes of diffusion signal models (DTM and SFM), assessing model-accuracy for predicting diffusion signals. A first contribution of this work is that we show that both models predict measurement in an independent data set more accurately than assuming the independent set will replicate the first data set (test-retest reliability). A second contribution is showing that the cross-validated model-accuracy of the DTM and SFM models are nearly identical at b-values of 1000, but at 2000 and above there is a small advantage to the SFM model-accuracy. This result is replicated in two different acquisition schemes on 6 different participants, and across several diffusion-weighting b-values.

There is consensus that DWI needs further validation, but there is no consensus about the appropriate validation methods. One approach is to compare fascicle models using ex-vivo measurements and phantoms [49,50]. A second is to compare models to simulations [28]. These are valuable analyses, but they do not address model-accuracy for any specific data set. The specificity is important because there is substantial variation in scanner hardware and subject populations. Hence, a pulse sequence that is optimal at one institution may be sub-optimal or even unfeasible at another institution or a different subject population. Because no single pulse sequence and processing pipeline will be optimal under all conditions, it is essential that researchers have methods to evaluate the method they use with respect to the data they collect. Such *in vivo* validation complements validation using ex vivo data, phantoms and simulations; they are not equivalent nor in conflict (see also [42]).

A third contribution of this work is that we provide a complete computational methodology for performing *in vivo* validation of diffusion models and processing pipelines. Using these methods, we show how to assess the effect of the number of directions and the b-value on model accuracy. We release the implementation of this methodology as open-source software.

## SFM and DTM model-accuracy

The SFM predicts the diffusion signal very well. The relative RMSE of a perfect model, assuming Gaussian measurement noise, is 0.707. The SFM model rRMSE is as low as 0.76, approaching the optimal level. The rRMSE of the DTM model is also not far from optimal, although it is inferior to the SFM model, particularly in regions of the brain with major fiber crossings and at high b-values (compare Figures 4 and 7).

There is a great deal of interest in making measurements at high b-values (4000 and up). The potential improvement in angular resolution, however, comes at the cost of reduced signal-to-noise ratio Specifically, in these data, the signal level drops while the noise level remains approximately constant (Figure 1). Because MRI data is non-negative, it has a Rician distribution, rather than a Gaussian distribution [51]. In principle, this should complicate model-fitting efforts, because the distributional assumptions do not necessarily hold. In practice, the signal level in the white matter is sufficiently high such that the noise is indistinguishable from a Gaussian distribution (Figure 2). In analyzing the tradeoff of signal-to-noise, we find that the model-accuracy of the SFM is better than that of the DTM at $b \geq 2000$ s/mm$^2$, but about the same or only slightly better at b = 1000. Additionally, as the b-value increases, the SFM fits improve but the DTM fits do not (Figures 6, 10).

While DTM model-accuracy is excellent at low b-values, it does not have good parameter-validity: The principal diffusion direction of the tensor does not match the fascicle orientations. Even so, the DTM can be useful as a coarse summary of white matter biology. The literature contains many examples where DTM parameters explain a substantial amount of variance in behavioral measures [18–20], showing the utility of the diffusion measurement itself in revealing a relationship between biology and mind.

Models of the SFM variety are a better basis for white matter tracking algorithms because the fascicle directions are estimated more accurately [2,42]. This is demonstrated in simulations using a small number of known crossing-angles (Figure 9). Nevertheless, further work is needed to determine the parameter-validity of the fODF estimates. For example, the fODF estimates are sensitive to assumptions about the fascicle response function [29] as well as the discretization of the model directions [52].

## The importance of cross-validation

Table 1 summarizes papers that evaluate diffusion models within the voxel. These methods differ from the approach presented here in several ways. First, most of these papers use simulations or exotic data sets that may not be comparable with respect to artifacts and noise obtained in typical human data sets with a clinical scanner. Hence, the results may not generalize to data collected at standard resolution and field strength. Second, most of these papers (with the exception of [8,25]) assess parameter-reliability or parameter-validity (relative to a model or phantom); they do not evaluate how well the model predicts the diffusion signal (model-accuracy). The work here is also the first assessment of model accuracy compared to test-retest reliability.

For the purpose of model comparison, the cross-validation approach is asymptotically equivalent to the Akaike Information Criterion (AIC, [53]). Cross-validation has the advantage that unlike the AIC it does not require an explicit calculation of the number of parameters. This calculation is not always straightforward, particularly when regularization is used is the model fitting procedure [54]. Cross-validation also has the advantage that unlike ANOVA it does not require nested models [25] or an assumption of Gaussian noise.

## SFM univariate summary measures

An attractive feature of the DTM is the associated univariate parameters that are derived from the tensor: fractional anisotropy (FA), radial and longitudinal diffusivity, and mean diffusivity. The SFM and similar models would benefit from having similar, informative, univariate summaries.

Dell'Acqua et al. proposed an interesting statistic using another SFM algorithm [55]. They define the Hindrance Modulated Orientation Anisotropy (HMOA) as the sum of the estimated fascicle weights in each voxel normalized so that a value of 1 is the highest possible value that can be realistically measured in a biological sample.

A second statistic is the number of distinct fascicles with a peak weight greater than some threshold [35,55]. Given that the number of fascicles may depend on the regularization constraints applied to the fODF and given that many regularization conditions produce equally accurate predictions of the signal (see S1 Figure), this statistic varies dramatically within a voxel, even for the same measurement, and we do not believe that it is very reliable.

Based on SFM, we can define two other statistics. The first is an SFM analog of FA: the fascicle-anisotropy (FA), which is the ratio of the sum of the fascicle weights, and the anisotropy weight $W_0$: FA = $\sum \beta / W_0$. A second useful univariate statistic measures an index of the distribution of angles between the fascicle directions, which we refer to as Dispersion Index (DI): $DI = \sum_{i=2}^{n} \frac{\beta_i^2 \sin(\alpha_i)}{\sum_i \beta_i^2}$, where $\alpha_i$ is the angle between the i[th] fascicle and the first fascicle, which is the one with the largest weight. This measure is larger when there are many large values of fascicle weights with large angles between them. The dispersion index summarizes the number of distinct crossing fiber populations within a voxel and is an interesting target for future research. Maps of FA and DI are shown in Figure 11.

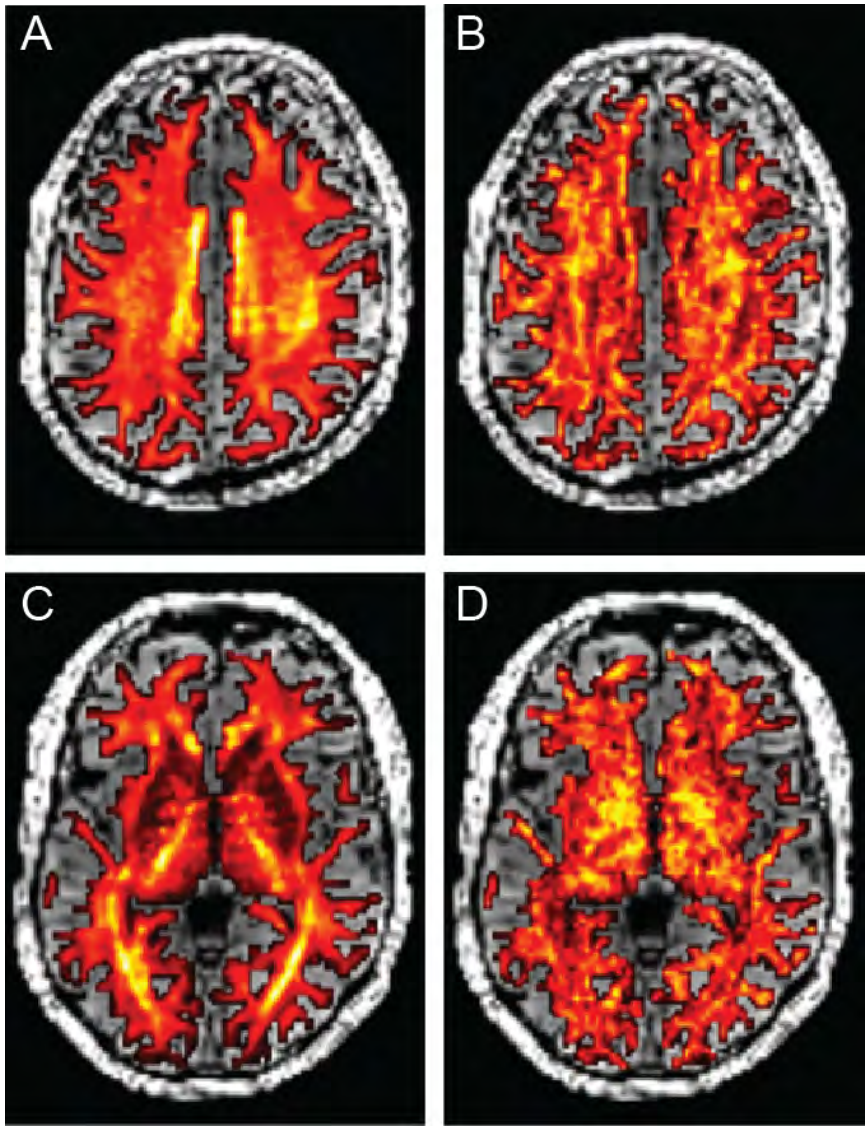

*Figure 11: Rotation-invariant statistics based on the SFM. The SFM leads to two rotation-invariant statistics that are calculated in every voxel, and shown here in axial sections at the height of the Centrum Semiovale (A, B; compare to Figures 4B, 7B) and the Optic Radiation (C, D; compare to Figures 4C, 7C). The Fiber Anisotropy (FA; A, C) is an indication of the total fiber fraction, relative to W0. The Dispersion Index (DI; B, D) is an indication of the degree to which different fascicles cross each other within each voxel.*

## Conclusion: Voxel-wise diffusion models are useful

For most of the white matter, but not all, DTM prediction error of a second data set is smaller than the error in test-retest reliability between the second and first data sets. The SFM model provides an even better fit to the data throughout the brain, and a substantially better fit in regions where the DTM does poorly. The DTM and SFM fits are accurate across b-values, although SFM is slightly more accurate at high b-values. At a b-value of 4000 s/mm$^2$ the DTM predicts an independent set of data more accurately

28than the test-retest prediction in 97% of the voxels; the SFM predicts the independent data set more accurately than test-retest reliability in 99.9% of the voxels. Because these models have high model-accuracy, replacing the data with the model prediction reduces noise. Hence, it is useful to perform subsequent analyses, such as tractography, using the model prediction rather than the raw measurements.

A major advantage of the SFM is that it provides a good orientation estimate when there are two fiber directions: SFM models explicitly estimate the fODF within each voxel [1,3,28]. An important alternative approach embodied in diffusion spectrum imaging (DSI, also called q-space imaging) does not make an explicit local model of the fODF [56,57]. The DSI approach is summarized as a "model-free diffusion MRI technique … without the need for a priori information or ad hoc models" ([56], pg. 1385). These methods measure diffusion signals in multiple directions and using multiple b-values. The complete set of data is used to derive a probability distribution that guides tractography, and there is no explicit commitment associating local extrema in the function with fascicles. As they are model-free, these methods do not discriminate between reliable signals, and irreproducible noise (Figure 8).

The cross-validation analysis and rRMSE measure described here show that having a model is useful for characterizing data with noise. Specifically, cross-validated model-accuracy is higher than test-retest reliability so that, the model reduces the measurement noise. This is true for the models that we have analyzed here, and could in principle be true for the many other available models that analyze diffusion MRI data, including models that use spherical harmonics [28], or a q-space approach to modeling the fODF [58,59], and models that use multiple tensors [60], or deconvolve the signal with stick-like functions [1]. All of these models predict the signal and their model-accuracy can be evaluated using the framework that we provide.

# Acknowledgments

Thanks to L.M. Perry for assistance in all stages of this study, to R.F. Dougherty for help with analysis and interpretation and to K. Chan for help with software implementation. Thanks to Q. Zhao and T. Hastie for useful discussions.# References

1. Behrens TEJ, Berg HJ, Jbabdi S, Rushworth MFS, Woolrich MW. Probabilistic diffusion tractography with multiple fibre orientations: What can we gain? Neuroimage. 2007;34: 144–55.

2. Tournier J-D, Calamante F, Connelly A. MRtrix: Diffusion tractography in crossing fiber regions. Int J Imaging Syst Technol. 2012;22: 53–66.

# Supporting information

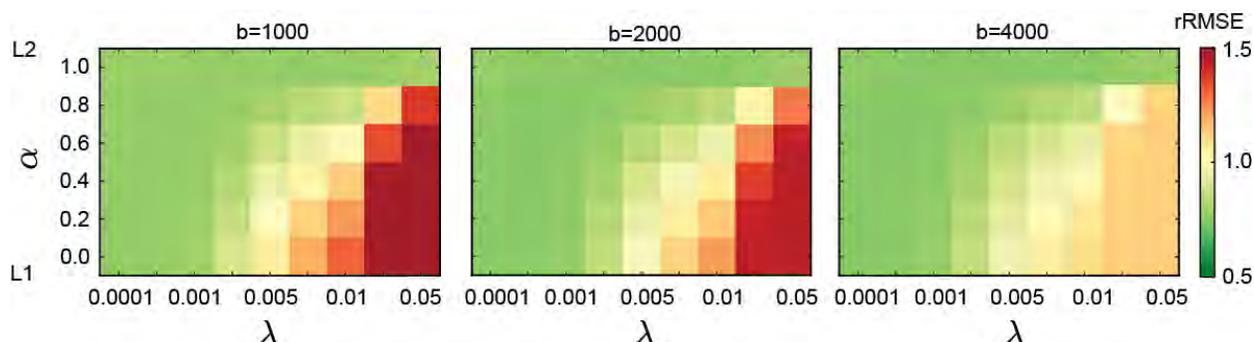

**S1 Figure: Regularization parameters for the SFM fit using Elastic Net.** We used regularization and cross-validation to (a) prefer solutions that minimize the number of fascicles and (b) prevent over-fitting. To find the appropriate setting of the regularization parameters **λ** and **α**, we used a cross-validation approach. The SFM was fit on one set of data for a range of **λ** and **α** values. For each combination the SFM was fit to one data set and the prediction error was calculated using the other data set. We choose **λ** and **α** that minimize the median *rRMSE* across white matter voxels. We explore the effects of regularization and the trade-off of different sets of constraints on the accuracy of the fit. The best setting of these parameters is to a relatively low degree of regularization ($\lambda = 0.0005$) and relatively L1-weighted constraint ($\alpha = 0.2$). These are the parameters used in all the SFM model fits.

# Table 1: previous approaches to model evaluation

| Publication | Experimental preparation | Instrument and conditions | Model evaluation method | Note |
|---|---|---|---|---|
| Alexander, Barker and Arridge, 2002 [25] | Human | 60 DW directions 1.7 x 1.7 mm inplane 2.5 mm throughplane Field strength not mentioned (assuming 1.5T) | F-tests comparing nested models of the fit | Model-accuracy |
| Jones, 2003 [9] | Human | Field strength= 1.5 T 64 | Confidence | Parameter- |
| | | directions resolution = 2.5 mm³ isotropic | interval in repeated estimates of the PDD of a diffusion tensor. | reliability |
| Tuch et al. 2004 [24] | Human | Field-strength=3T Resolution = 3.125 x 3.125 x 3.1 mm³ 126 directions b-value =1077 | RMSE of the DTM signal relative to the measured signal (Equation 5) | Figure 5 in this paper shows that the DTM does not model the signal well in voxels in which the signal is oblate (an estimate of model-accuracy) |
| Chang et al. 2005[61] | Simulation | | Accurate recovery of the simulated FA | Parameter-validity |
| Chung et al. 2006 [11] | Simulation | | Accurate estimation of the variability of DTM-derived measures | Parameter-validity |
| Koay et al. 2006 [38] | Simulation | | Chi-square distribution and error in estimating the trace of a simulated tensor sigal | Parameter-validity |

| Whitcher et al. 2008 [12] | Simulations and human DTI measurements | 3T | Reliability of DTI-derived measures assessed by means of the wild bootstrap. | Parameter-reliability |
|---|---|---|---|---|
| Tournier et al. 2008 [50] | Phantom, 20 and 90 um fused silica tubing | 9.4T spectrometer 25/32 mm inplane 3.6 mm through-plane b-value=8000 | Phantom direction recovery. | Parameter-validity |
| Panagiotaki et al. 2012 [8] | Rat corpus callosum | 9.4T small bore spectrometer 2mm/256 (80 microns inplane) thickness 500 microns 5 directions | Bayesian Information Criterion of fits across many pulse sequence types | Model-accuracy |

**Table 2: data-sets analyzed**

| Diffusion weighting (b-value) | Spatial resolution | # diffusion directions | # subjects | URL for download | Duration |
|---|---|---|---|---|---|
| 1000 | 2 x 2 x 2 mm$^3$ | 150 | 2 | http://purl.stanford.edu/ng782rw8378 | 19:03 |
| 2000 | 2 x 2 x 2 mm$^3$ | 150 | 2 | http://purl.stanford.edu/ng782rw8378 | 20:56 |
| 4000 | 2 x 2 x 2 mm$^3$ | 150 | 2 | http://purl.stanford.edu/ng782rw8378 | 22:53 |
| 2000 | 1.5 x 1.5 x 1.5 mm$^3$ | 96 | 6 (this includes the 2 subjects in the other data sets) | http://purl.stanford.edu/rt034xr8593 | 39:03 |